\documentclass[epj]{svjour}  
\input epsf.sty    
\usepackage{graphics}
\begin{document} 
\title{Electromagnetic structure of the deuteron}
\subtitle{review of recent theoretical  and experimental results}
\author{Franz Gross
}                     
%
%
\institute{Jefferson Laboratory, 12000 Jefferson Avenue,
Newport News, VA 23606; \email{gross@jlab.org}
}
\date{Received: September 13,2002}
%
\abstract{Recent high energy measurements of elastic $ed$ scattering
support the use of a relativistic theory based on an accurate description
of the $NN$ channel, but theory needed for an
understanding of the high energy deuteron photodisintegration cross
sections and polarization observables is not yet mature.    
} 
\maketitle
\vspace*{-3.5in} 
\leftline{JLAB-THY-02-43}
\vspace*{3.5in} 


\section{Introduction}
\label{intro}
This talk reviews recent theoretical and experimental results for  
elastic electron deuteron scattering (yielding the deuteron form
factors), threshold electrodisintegration ($e+d\to
e'+p+n$ where the mass of the final $np$ pair, $W$, is only a few MeV
above the threshold value of $m_p+m_n$), and high energy deuteron
photodisintegration ($\gamma+d\to p+n$).  The talk is based on the
complete reviews of Refs.~\cite{GVO,S,GG}, with a few new
results not previously reported.  


\section{Deuteron Wave Functions}

\begin{figure}
\begin{center}
\mbox{
   \epsfxsize=2.8in
\epsffile{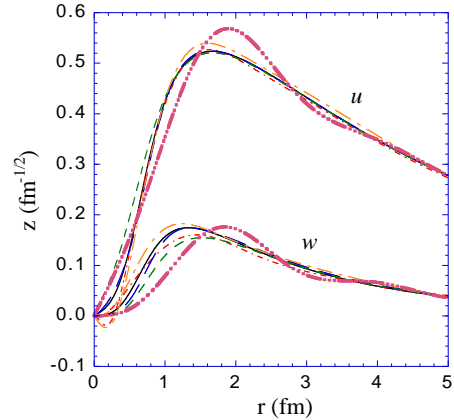} 
}
\end{center}
\vspace*{-0.2in}
\caption{The $u$ and $w$ wavefunctions in $r$ space 
for six models discussed in the text.}
\label{fig:1}
\end{figure}

\begin{figure}
\begin{center}
\mbox{
   \epsfxsize=2.8in
\epsffile{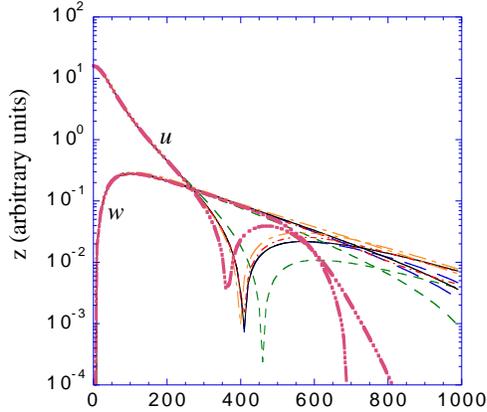} 
}
\end{center}
\caption{The $u$ and $w$ wavefunctions in $p$ space 
for six models discussed in the text.  (Each model is normalized so
that $u(0)=16$.)}
\label{fig:2}
\end{figure}

\begin{figure}
\begin{center}
\mbox{
   \epsfxsize=2.8in
\epsffile{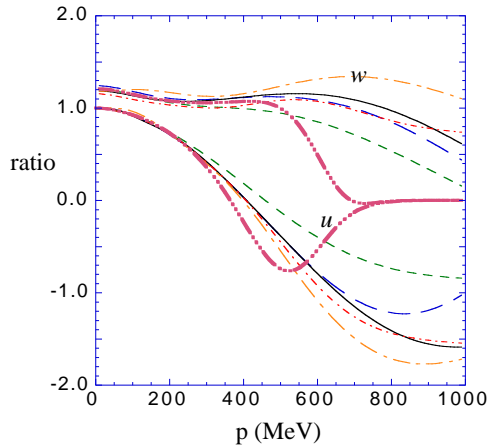} 
}
\end{center}
\caption{The scaled $u$ and $w$ wavefunctions in $p$ space 
for six models discussed in the text.  (The scaling function used
here is defined in Ref.~\cite{GG}.)}
\label{fig:3}
\end{figure}

The deuteron wave functions are calculated from a potential (or a
relativistic kernel) that has been fitted to $NN$
scattering data below lab kinetic energies of 350 MeV.  Figs.
\ref{fig:1},\ref{fig:2}, and \ref{fig:3} show the coordinate and
momentum space wave functions for six models: Argonne AV18
\cite{AV18}, Paris \cite{Paris}, CD Bonn \cite{CDBonn},  IIB
\cite{gvoh92}, W16 \cite{sg97} and the recent Idaho potential
\cite{Idaho}.  These figures are modified versions of those
in Ref.~\cite{GG}; the principle change is the inclusion of the new
Idaho wave functions (the heavy dashed-triple dotted line) which have
a rapid but smooth cutoff above 600 MeV momentum.  This cutoff
explains the ripples in the $r$ space wave functions, and will have a
profound effect on the deuteron form factors.    

\section{Deuteron Form Factors}
\label{sec:1}

The deuteron form factors are defined by the relativistic deuteron
current, which has the form
\begin{eqnarray}
\left<d'|J^\mu|d\right>&=&-G_1(Q^2)\;
[\xi'^*\cdot\xi]\,(d^\mu+d'^\mu) \nonumber\\
&&+G_3(Q^2)\frac{(\xi'^*\cdot q)(\xi\cdot
q)}{2m_d^2}
\,(d^\mu+d'^\mu)  \nonumber\\
 &&- G_M(Q^2)\;[\xi^\mu(\xi'^*\cdot q)
-\xi'^{*\mu}(\xi\cdot q)]\, ,\label{deutcurrent}
\end{eqnarray} 
where $Q^2=-q^2$, with $q=d'-d$, is the square of the four-momentum
transferred by the electron, and $d(d')$ and $\xi(\xi')$ are the
incoming (outgoing) deuteron four momenta and polarization,
respectively.  Instead of $G_1$ and $G_3$, it is customary to use the
linear combinations  
\begin{eqnarray}
G_C&&=G_1 +\frac{2}{3}\,\eta\,G_Q\nonumber\\ 
G_Q&&=G_1-G_M+(1+\eta)G_3\, ,
\end{eqnarray}
with $\eta=Q^2/4m_d^2$.  At $Q^2=0$, the form factors $G_C$, $G_M$, 
and $G_Q$ give the charge, magnetic and quadrupole moments of the
deuteron
\begin{eqnarray}
G_C(0) &= 1 &\qquad ({\rm in\ units\ of}\, e )\nonumber\\ 
G_Q(0) &= Q_d &\qquad ({\rm in\ units\ of}\, e/m_d^2)\nonumber\\
G_M(0) &=\mu_d &\qquad ({\rm in\ units\ of}\, e/2m_d) \, .     
\end{eqnarray}
The form factors must be extracted from measurements of the elastic
$e+d\to e'+d'$ cross section, and the polarization transfer $T_{20}$,
which give the combinations
\begin{eqnarray}
&&A(Q^2) = G_C^2(Q^2) + {{8}\over{9}} \eta^2 G_Q^2(Q^2) + 
 {{2}\over{3}} \eta G_M^2(Q^2)\nonumber\\
&&B(Q^2) = {{4}\over{3}} \eta(1+\eta) G_M^2(Q^2)\nonumber\\
&&\tilde{T}_{20}= - \sqrt{2}\; { {y(2+y)}\over{1 + 2 y^2} } \qquad
{\rm with}\;\; y = 2 \eta G_Q / 3 G_C
\, .
\label{AandB}
\end{eqnarray}

\subsection{Nonrelativistic Theory} 

\begin{figure}
\begin{center}
\mbox{\vspace{-0,3in}
   \epsfxsize=3.0in
\epsffile{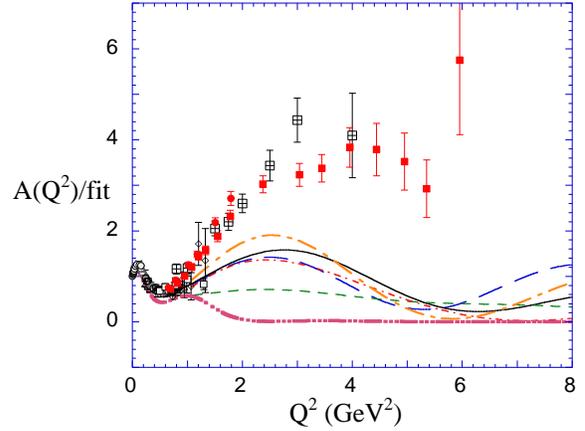} 
}
\end{center}
\caption{The nonrelativistic inpulse approximations predictions for
the six models compared to data.  (The predictions and data are all
divided by the scaling function given in Ref.~\cite{GG}.) }
\label{fig:4}
\end{figure}

Figure \ref{fig:4} gives a comparison of the $A$ structure function
with the theoretical predictions of the nonrelativistic impulse
approximation (the NRIA includes {\it no relativistic effects or
exchange current contributions\/}).  The figure shows that the
NRIA fails at large $Q^2$ by a factor of 4 to 8 for most of the models,
and by a much larer factor for the Idaho model.  Because the deuteron
form factors are purely isoscalar, and isoscalar exchange currents are
largely of relativistic origin, an explanation of the form factors
will require a relativistic theory.  The figure also shows that the
Idaho model in NRIA, with no high momentum components, greatly
underpredicts the $A$ structure function at high $Q^2$, requiring
exceptionally large exchange currents in order to explain the data. 

\subsection{Relativistic Theory}

\begin{figure*}
\begin{center}
\mbox{
   \epsfxsize=4.8in
\epsffile{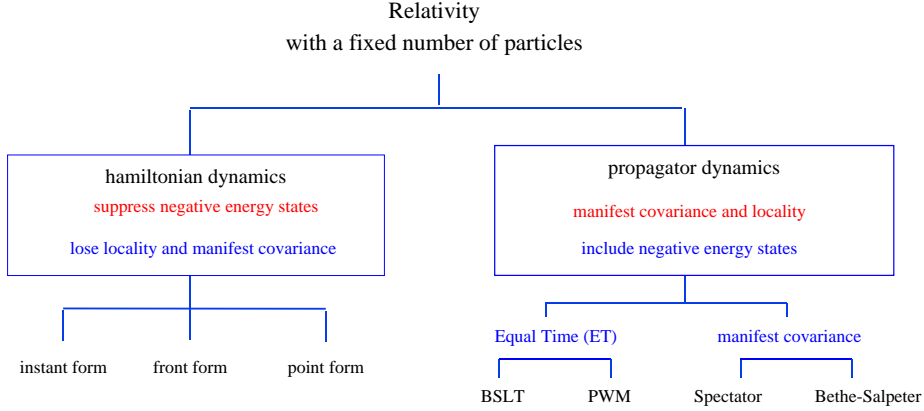} 
}
\end{center}
\caption{Classification of relativistic methods.  The class on the right
is propagator dynamics (${\cal P}$) and on the left is hamiltonian
dynamics (${\cal H}$). }
\label{fig:5}
\end{figure*}

The deuteron form factors are an ideal system for a test
relativistic theory.  The data is accurate, and the form factors
are reasonably well approximated by the nonrelativistic theory of a
single $NN$ channel, making them comparatively simple to calculate.  
Many groups have accepted the challenge and nearly all of the possible
relativistic approaches outlined in Fig.~\ref{fig:5} have been
tried.  The methods break into two large classes, which I call
propagator dynamics (based on field theory) and hamiltonian dynamics
(based on the three choices of dynamics originally classified by
Dirac).  The predictions of seven different relativistic
calculations, listed in Table \ref{table:1}, are compared in
Ref.~\cite{GG} and in the left panel of Fig.~\ref{fig:8} below.  As shown
in the table, two of these use propagator dynamics, and five use
hamiltonian dynamics.   Space does not permit a discussion of these
approaches here; look at Ref.~\cite{GG} for a detailed discussion. 
Here I want to highlight some features of the VOG approach
\cite{vdg95}, which uses the manifestly covariant spectator theory
\cite{GCS}.

In this theory the $NN$ propagator is obtained from the covariant
Bethe-Salpeter propagator by replacing one of the nucleon propagators by
\begin{eqnarray}
S(p)=\frac{m+\not p}{m^2-p^2}\to2\pi\delta_+(m^2-p^2)\!\sum_s u({\bf
p},s)
\,\bar u({\bf p},s)\, .\quad
\end{eqnarray}
This substitution insures that the nucleon with momentum $p$ is on its
positive energy mass shell, reducing the 4-d Bethe-Salpeter
equation to the covariant 3-d spectator equation.  The resulting
relativistic wave functions depend only on the three-momentum, as in
the nonrelativistic case, but can be Lorentz transformed from
frame-to-frame using transformations that do not depend on the details of
the dynamics ({\it i.e.\/} they are kinematic).  The boost transformation
is  
\begin{eqnarray}
{\cal B}(\Lambda)\Psi^\mu_{\alpha\lambda}(p,d)\,\xi_\mu
=B_{\alpha\alpha'} \Psi^\mu_{\alpha'\lambda'}(\Lambda p,\Lambda d) 
\,(\Lambda\xi)_\mu\, d^{(1/2)}_{\lambda'\lambda}(\omega)\, ,\quad
\end{eqnarray}
where $\alpha$ is the Dirac index of the off-shell particle, $\lambda$
the helicity of the on-shell particle, and $\omega$ is the Wigner rotation
angle for the helicity of the on-shell particle.  This exact formula makes
it possible to calculate the recoil of the outgoing deuteron exactly to
all orders in $(v/c)$, and is used in calculations of the form factors. 
Finally, I emphasize that the close connection of this formalism to field
theory makes it possible to design currents that are complete, physical,
and consistent. 

\begin{table}
\caption{Features of seven relativistic models.}
\label{table:1}
\begin{tabular}{lclcc}
model & class & description &in & complete\\
&&&Fig.~\ref{fig:6}& current \\
\hline\noalign{\smallskip}
VOG \cite{vdg95} & ${\cal P}$ & Spectator  & yes &yes\\  
PWM \cite{PW} & ${\cal P}$ & modified & no & no \\&&
Mandelzweig-Wallace \\
FSR \cite{FS01} & ${\cal H}$ & instant-form;& no & yes \\&&no $v/c$
expansion \\ 
ARW \cite{ARW00} & ${\cal H}$ & instant-form;  & yes & yes\\&&with $v/c$
expansion \\
CK \cite{CK99} & ${\cal H}$ & front-form; & no  & no  \\&&dynamical
light-front  \\ 
LPS \cite{LPS00} & ${\cal H}$ & front-form;& yes & no  \\
&&fixed light-front \\
AKP \cite{AKP01} & ${\cal H}$ & point-form & yes  & no\\ 
\hline\noalign{\smallskip} 
\end{tabular}
\end{table}

\begin{figure}
\begin{center}
\vspace{-0.2in}
\mbox{
   \epsfxsize=3.0in
\epsffile{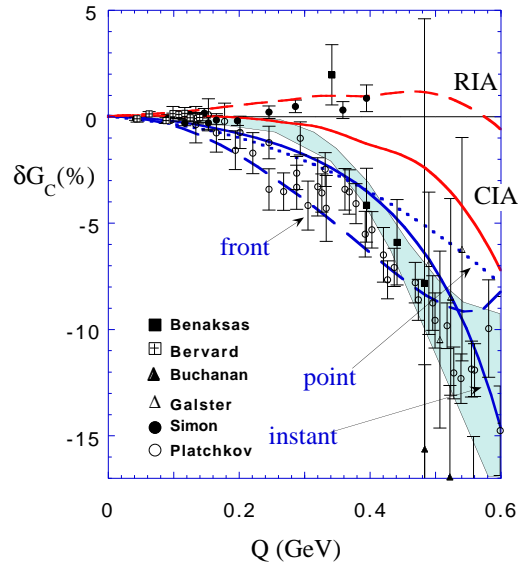} 
}
\end{center}
\vspace*{-0.1in}
\caption{Comparison of selected relativistic calculations of $G_C$ with
charge form factor data.  The
shaded area is the best fit (with errors) determined by Sick \cite{S}. 
All curves and data are expressed as a percentage difference
from the AV18 calculation in NRIA. }
\label{fig:6}
\end{figure}     

\begin{figure*}
\begin{center}
\mbox{
   \epsfxsize=5.0in
\epsffile{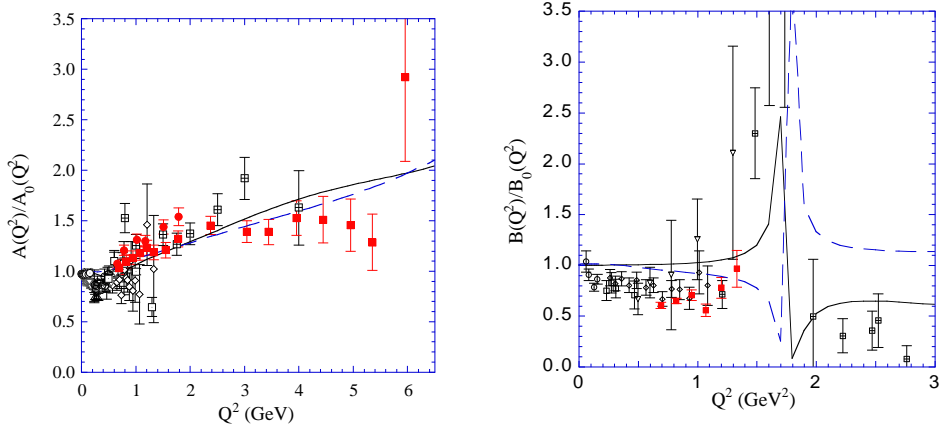} 
}
\end{center}
\caption{Ratios of the structure functions $A$ and $B$ with different
choices for the form factors $F_3$ or $f_{\rho\pi\gamma}$ normalized to
$A_0$ and $B_0$ calculated with the ``standard'' choices $F_3$=dipole and
$f_{\rho\pi\gamma}=0$.  Solid line: $F_3$=tripole,
$f_{\rho\pi\gamma}=0$.  Dashed line: $F_3$=dipole,
$f_{\rho\pi\gamma}$=dipole with $\Lambda_d^2$=1.5 GeV$^2$.  }
\label{fig:7}
\end{figure*}     

Figure \ref{fig:6} shows how experimental values of $G_C$ extracted from
measurements of $A$ (corrected by removing the magnetic and quadrupole
contributions\footnote{These corrections, of the order of 5\% at the
larger $Q$ shown, were omitted from Ref.~\cite{GG}.}) compare to
Sick's best fit \cite{S}, and to different relativistic
calculations.  Using the Argonne AV18 NRIA as a nonrelativistic standard,
the data  support the conclusion that relativistic effects grow rapidly
for $Q\geq0.3$ GeV.  Furthermore, the accuracy of the measurements is
comparable to the differences between relativistic approaches,
suggesting that a  precision 1\% measurement of $G_C$
would further constrain relativistic
theory \cite{JLabexpt}.  The apparent inconsistency
between the Simon and Platchkov measurements is an additional motivation,
but Sick has studied both measurements in detail and reports that the
data sets are consistent within statistical and systematic errors
\cite{S2}.

I will conclude this discussion by calling attention to one aspect of the
VOG spectator calculation of special theoretical interest.  Feynman showed
years ago that a photon amplitude will be gauge invariant if (i) the
coupling of the photon to the (off-shell) constituents inside the
amplitude satisfies the Ward-Takahashi identities, and (ii)
electromagnetic couplings to all particles inside the amplitude are
included.  In 1987, Riska and I found out how to generalize this result
to calculations using either Bethe-Salpeter amplitudes or spectator
amplitudes involving {\it composite\/} constituents \cite{GR87}.  In the
VOG theory, off-shell nucleons are described by a dressed propagator
$S(p)=h^2\,S_0(p)$, where
$h=h(p)$ is a scalar function of the square of the off-shell nucleon four-momentum
$p^2$ [with its only parameter fixed by fits to the $NN$ data], and
$S_0$ is the undresed nucleon propagator.  Use of this propagator
requires that the {\it simplest, but not unique\/} single nucleon current
must have the form

\begin{eqnarray}
 j^\mu(p',p) &=&
f_0(p',p)\left(F_1(Q^2)\,\gamma^\mu+{F_2(Q^2)\over2m}
i\sigma^{\mu\nu}q_\nu\right)\nonumber\\&&+
g_0(p',p)\,F_3(Q^2){m-\not\!p'\over2m}\gamma^\mu
{m-\not\!p\over2m} \, , \quad \label{onej}
\end{eqnarray}
where $f_0$ and $g_0$ are known functionals of $h$ and
$h'=h(p')$ [with
$p'$ the four momentum of the outgoing nucleon], and $F_3(0)=1$ but is
otherwise undefined.  In the original calculations, $F_3$ was taken to
be a dipole, but if it is made somewhat harder, {\it all\/} of the
deuteron observables can be well described by the spectator
theory.  This remarkable result is shown in Fig.~\ref{fig:7}. 
Neglecting the $\rho\pi\gamma$ exchange current, and choosing
$F_3$ to be the tripole $(1+Q^2/\Lambda_t^2)^{-3}$ with $\Lambda_t^2=$ 5
GeV$^2$, brings {\it all\/} of the observables into excellent agreement
with the high $Q^2$ data measured at SLAC and JLab (including $T_{20}$
not shown).  Alternatively, using the ``standard'' dipole form,
$F_3=(1+Q^2/\Lambda_d^2)^{-2}$ with
$\Lambda_d^2=0.71$ GeV$^2$ and choosing a dipole form for the
$\rho\pi\gamma$ form factor (with $\Lambda_d^2=$ 1.5 GeV$^2$) will give
agreement for the structure function
$A$, but increases the disagreement in $B$.  I conclude that a reasonable
choice of $F_3$ will bring the VOG model into perfect agreement with the
deuteron data.  Of course, it remains to see if this choice will also
describe other electromagnetic processes, and this brings us nicely to
the next topic.

\begin{figure*}
\leftline{
\mbox{
   \epsfysize=4.3in
\epsffile{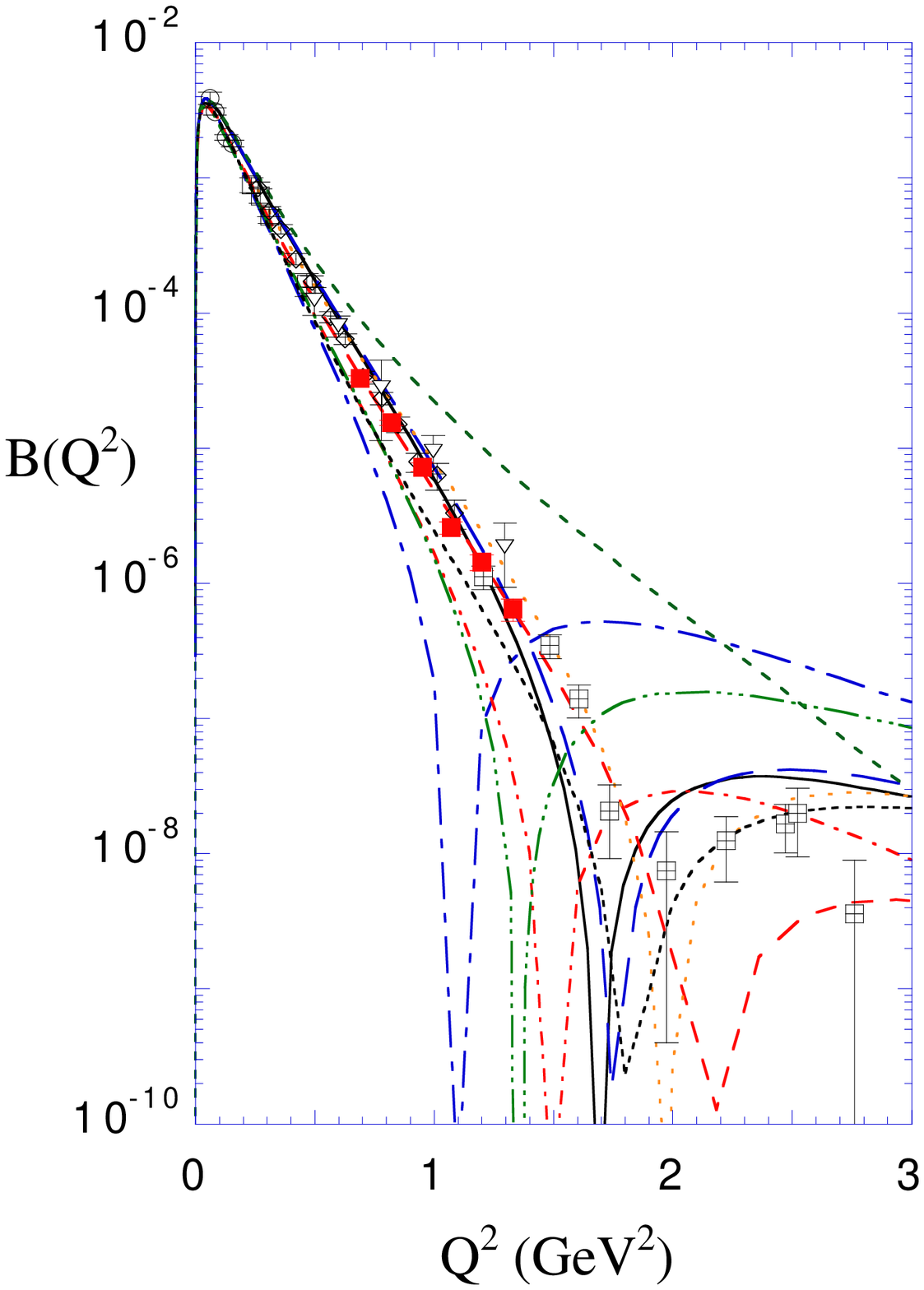} 
}}
\vspace*{-4.1in}
\hspace*{-0.3in}
\rightline{
\mbox{
   \epsfysize=3.8in
\epsffile{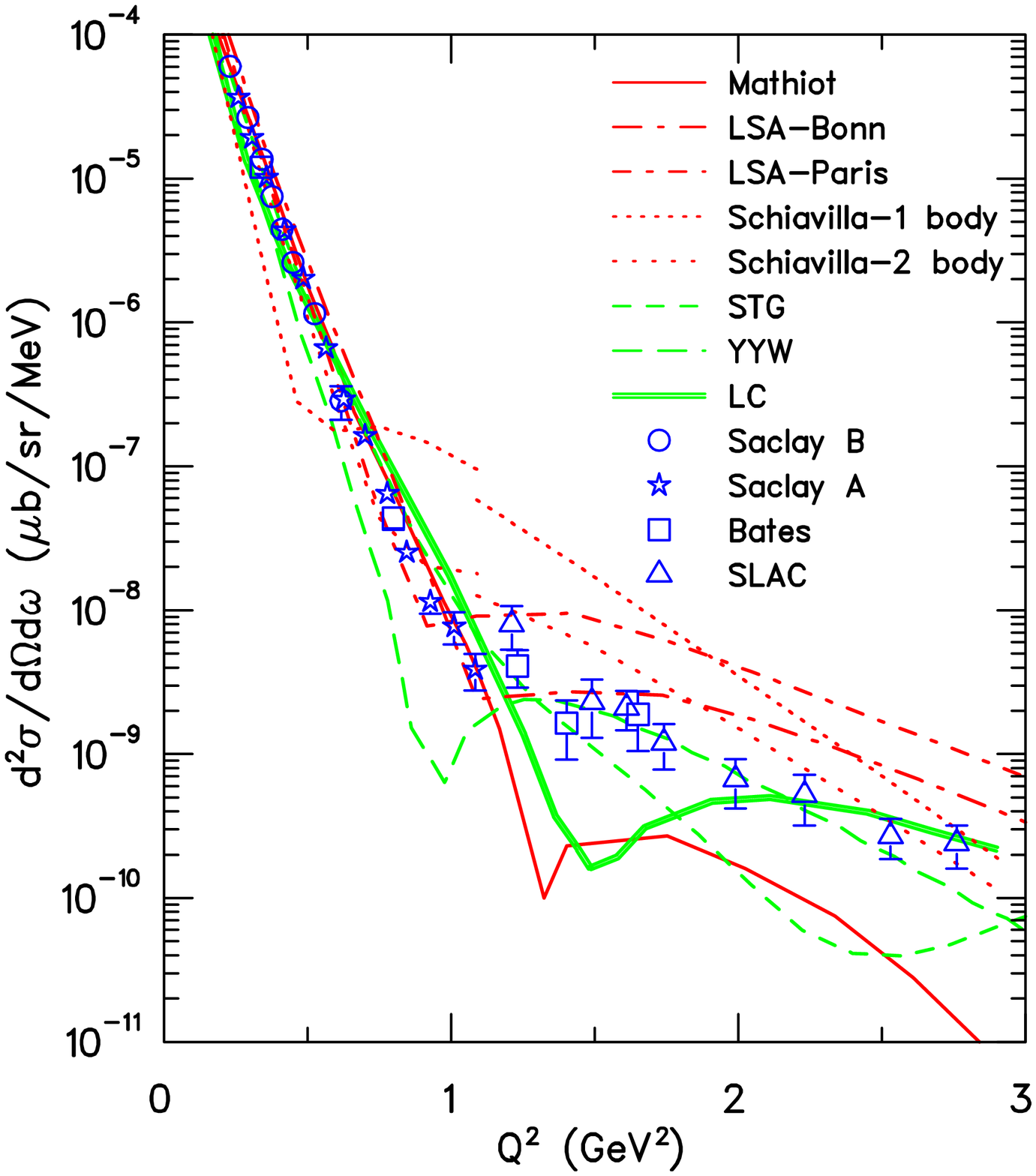} 
}}
\vspace*{0.1in}
\caption{Theory and data for the elastic $B$ structure function (left
panel) and the threshold electrodisintegration cross section (right
panel).  The calculations listed in Table \ref{table:1} are shown in the
left panel.}
\label{fig:8}
\end{figure*}     

\section{Threshold Electrodisintegration}

Deuteron electrodisintegration to an $np$ final state
with an invariant mass $W$ only a few MeV above the threshold,
$W_t=m_p+m_n$, is very closely related theoretically to elastic $ed$
scattering.  Here the final scattering state can be as reliably calculated
as the initial deuteron state using the same theory.  If the mass
of the final state is close enough to threshold, the transition will be
dominated by (purely magnetic) transitions to the $^1S_0$ final state
(transitions to the $^3S_1$ final state are suppressed by the
orthogonality of the states), and the transition amplitude will be purely
isovector.  Together with elastic scattering (purely isoscalar) the two
processes can be used to independently determine the isoscalar
and isovector exchange currents.

Figure \ref{fig:8} compares calculations of threshold
electrodisintegration with recent calculations of the $B$ structure
function (the magnetic counterpart of the electrodisintegration
amplitude).  In spite of the close theoretical connection between
these two reactions, few theoretical groups have calculated both.  Both
are very sensitive to the details of the theory, and together will
provide a stringent test.  More accurate data is needed for both
reactions, particularly in the region of the minima near
$Q^2\simeq$1.5 GeV$^2$.

\section{Photodisintegration}

There are significant differences between elastic $ed$
scattering using electrons of several GeV, discussed above, and
photodisintegration by photons of several GeV.  In the elastic reaction,
the final state remains bound, all of the virtual photon energy going
into the recoil of the final deuteron.  In the VOG relativistic
approach where the boost operators are purely kinematic, this recoil can
be calculated exactly, and one can hope to describe the process using
dynamics based on low energy physics.  In photodisintegration, however,
the bulk of the photon energy goes into exciting the final state, and 4
GeV photons have sufficient energy to excite all of the nucleon resonances
listed in the particle data book!  A hadronic calculation of this final
state is necessarily a complicated problem involving the coupling of
over 200 channels.  Clearly the theory of high energy
photodisintegration must either be based on an average treatment of
many hadronic channels, or on the use of quark degrees of
freedom, which provides the high energy alternative to hadronic based
descriptions. 

\begin{figure}
\begin{center}
\mbox{
   \epsfxsize=2.9in
\epsffile{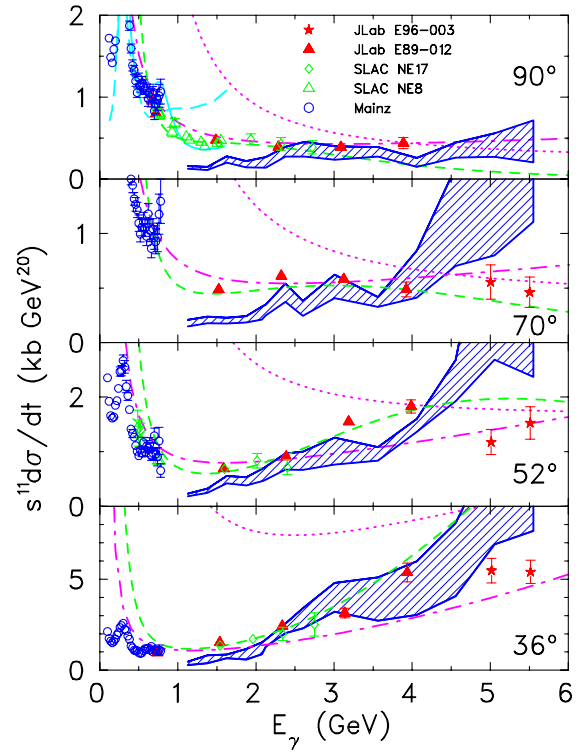} 
}
\end{center}
\caption{Photodisintegration cross section
$s^{11}d\sigma/dt$ versus incident lab photon energy for various fixed
angles.  For discussion and references see Ref.~\cite{GG}. }
\label{fig:9}
\end{figure}     

A number of theoretical descriptions using quarks or
of Reggie trajectories (a form of averaging over hadronic degrees of
freedom) have been applied to the description of photodisintegration. 
For a discussion see the review \cite{GG} or the talk by Patrizia Rossi
elsewhere in these procedings \cite{PR}.  Here we call attention only to
the status of the prediction from perturbative QCD, that the cross
section should fall like $s^{-11}$.  Figure \ref{fig:9} shows that
recent high energy data from JLab \cite{bochna98,schulte01} generally
confirm this behavor, but a model by Raydushkin (the dot-dashed line)
with an
$s^{-10}$ behavior does just as well.

\begin{figure}
\begin{center}
\mbox{
   \epsfxsize=2.8in
\epsffile{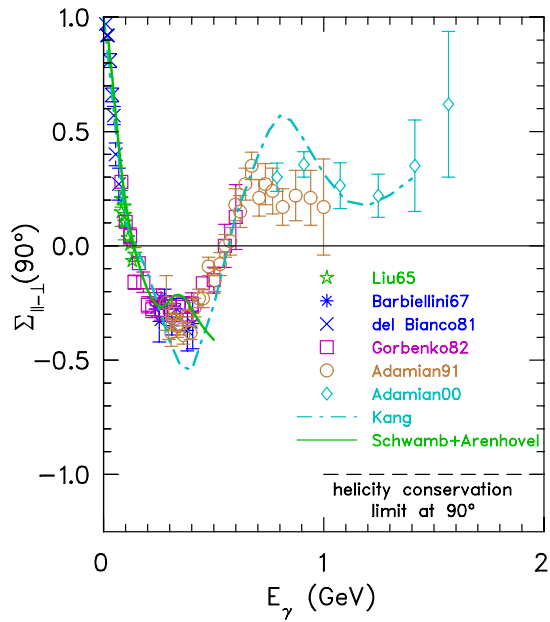} 
}
\end{center}
\caption{Polarized photon asymmetry $\Sigma$  at $\theta_{\rm cm}$
$=$ 90$^\circ$.  For discussion and references see Ref.~\cite{GG}. }
\label{fig:10}
\end{figure}

Polarization observables provide the most stringet test of any model, and
pQCD in particular.  Figure \ref{fig:10} shows recent high energy Yerevan
data \cite{adamian00} on the polarized photon asymmetry $\Sigma$.  The
result expected from pQCD (supplemented by additional assumptions) that
$\Sigma(90^\circ)\to-1$ as
$E_\gamma\to\infty$ is not confirmed.

\section{Conclusions}

Elastic scattering and photodisintegration with 4 GeV electrons are
closely related experimentally, but require very different theoretical
approaches.  For example, if reasonable adjustments are made in the high
$Q^2$ behavior of the current of an off-shell nucleon, elastic scattering
is well described by a relativistic theory based on low energy $NN$
scattering.  This study confirms our fundamental understanding of low
energy $NN$ scattering, shows us that the relativistic theory is under
control, and fixes the high $Q^2$ behavior of the off-shell nucleon
current.

A similar understanding of high energy photodisintegration is
not yet available; quark degrees of freedom, hidden in elastic
scattering, may eventually be required for an understanding of
photodisintegration.  However, the early successes of pQCD do not seem
sufficiently robust to serve as a basis for a deeper understanding in this
energy range.

\begin{acknowledgement}

It is a pleasure to thank Ron Gilman for the collaborative effort leading
to Ref.~\cite{GG}.  He supplied many of the figures in this minireview. 
I also thank Ingo Sick for helpful discussions and for a computer version
of his form factor fits, and Ruprecht Machleidt for numerical
values of his Idaho wave functions.  My apologies to the many people
who supplied information or whose work is not cited in this short
summary; they are listed and cited in \cite{GG}.  This work was supported
in part by the US  Department of Energy. The Southeastern  Universities
Research Association (SURA) operates the Thomas Jefferson National
Accelerator Facility under DOE contract DE-AC05-84ER40150.  It also is a
pleasure to acknowledge support under DOE grant No.~DE-FG02-97ER41032.

\end{acknowledgement}

\end{document}